# Urban water networks as an alternative source for district heating and emergency heat-wave cooling

Xiaofeng Guo[1,2*], Martin Hendel[1,2]

[1] ESIEE Paris, Université Paris Est, 2 boulevard Blaise Pascal - Cité Descartes, F-93162, Noisy Le Grand, France
[2] Univ Paris Diderot, Sorbonne Paris Cité, LIED, UMR 8236, CNRS, F-75013, Paris, France

## Abstract

Urban water networks can contribute to the energy transition of cities by serving as alternatives sources for heating and cooling. Indeed, the thermal energy potential of the urban water cycle is considerable. Paris is taken as an example to present an assessment of the field performance of a district-scale waste water heat recovery system and to explore potential techniques for emergency cold recovery from drinking or non-potable water networks in response to heat-waves. The case heat recovery system was found to provide significant greenhouse gas emission reductions (up to 75%) and limited primary energy savings (around 30%). These limited savings are found to be mainly due to the performance of the heat pump system. Three emergency cold recovery techniques are presented as a response to heat-waves: subway station cooling, ice production for individual cooling, and "heat-wave shelter" cooling in association with pavement-watering. The cold generation potential of each approach is assessed with a special consideration for mains water temperature sanitary limitations. Finally, technical obstacles and perspectives are discussed.

## 1 Introduction

Concentrating 60% to 80% of the world's energy consumption [1], cities are at the heart of the energy transition challenge facing humanity over the 21st Century. This challenge is made more difficult by the changes in climate expected over the course of the current century, which will gradually and inevitably affect the way energy is used to heat and cool buildings.

As climate change continues, cities will witness a decrease in their heating demand and an increase in their cooling demand. While the decrease in Heating Degree Days (HDD) forebodes energy savings, these may likely be compensated by the sharp increase in cooling demand [2]. This trend can be observed in many major cities across the globe and present a major challenge for the world's successful energy transition [3]. In Paris, as can be seen in **Figure 1**a), building energy demand is clearly heating-dominated. This is reflected by its average 2 352 ˚C.day of HDD, while cooling demand remains small with a total 17 °C.day of cooling degree days (CDD) (the threshold values used are 18 °C for heating and 24 °C for cooling) [4]. At the end of this century, climate change is expected to decrease HDD by 30%

* *Corresponding author: Xiaofeng Guo, xiaofeng.guo@esiee.fr, Tel: +33 14592 6058, Fax: +33 14592 6699 address: ESIEE Paris, department SEN, 2 bd Blaise Pascal BP99, 93162, Noisy Le Grand Cedex, FRANCE*

to 1 622 °C.day, while CDD should increase seven-fold to 127 °C.day [4]. This shift is already visible over the last few decades [5].

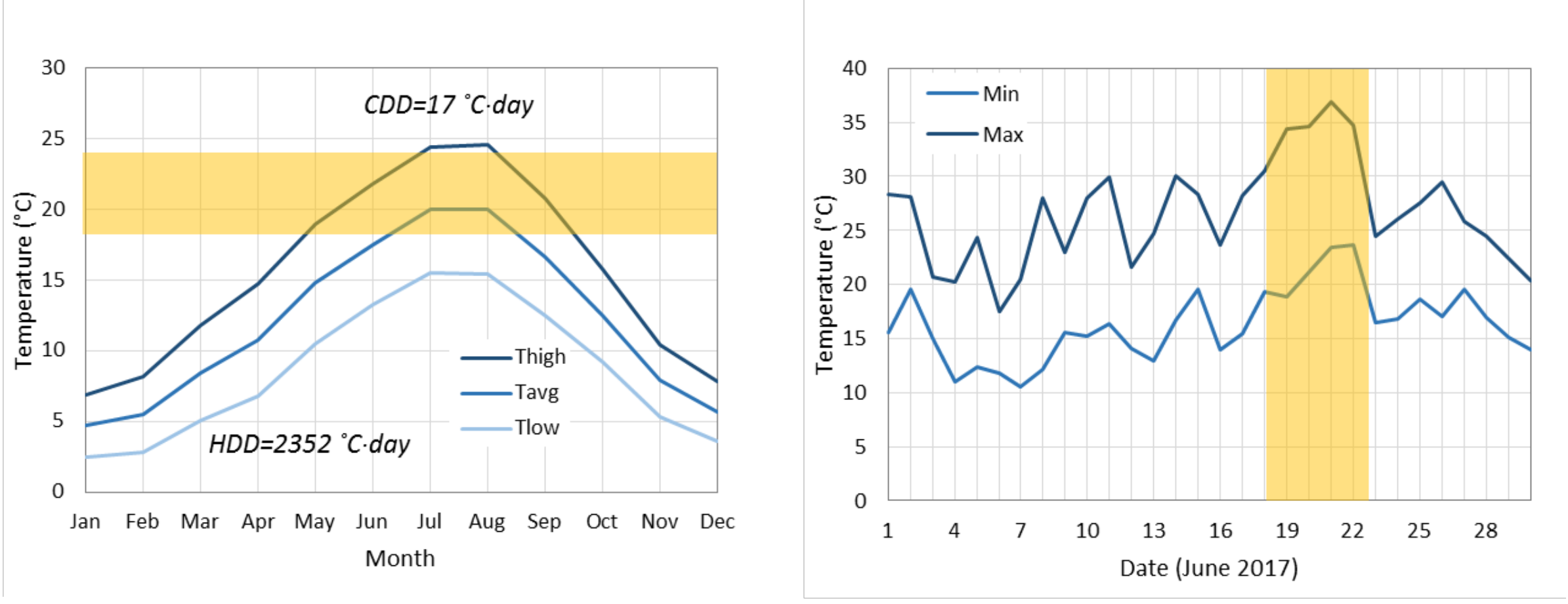

a) Monthly temperatures for Paris

b) Maximum and minimum daily temperatures during the June 2017 heat-wave

**Figure 1.** Air temperature data at the Montsouris weather station in Paris
The yellow band shown in the left gives the human comfort zone between 18 °C and 24 °C; The vertical band on the right outlines the peak of June 2017 heat-wave.

In addition to the local climate, which is the main determinant for building heating and cooling demand, cities are also subject to the urban heat island (UHI) effect. This localized warming phenomenon is the result of a combination of radiative trapping, increased heat storage, wind obstruction, low vegetation presence, low surface permeability and high concentrations of human activity along with corresponding heat release [6]. One should also mention the increase of individual, air source air-conditioning systems that intensify the UHI. These mechanisms cause higher air and surface temperatures in city centres relative to the surrounding rural areas, in the order of 1° to 3°C [7]. In terms of its impact on energy consumption, UHI tends to increase cooling demand and reduce that of heating.

Parallel to the global climate shift and UHI effect, the frequency of extreme weather events, in particular heat-waves, is expected to increase [8]. In Paris, heat-waves are expected to increase from 1 heat-wave day per year to as many as 26 days per year [4]. Combined with the UHI effect, these events pose a serious public health concern, as witnessed during the 2003 heat-wave [9]. Although infrequent, such events are characterized by high temperatures during more than 3 consecutive days, as shown in **Figure 1**b), and merit active cooling techniques. How to deal with short but intense emergency cooling needs in traditionally heating dominated regions is more of a public security issue than an energy efficiency concern.

In recent years, energy recovery from urban water networks has gained increasing attention from urban planners as well as water utility companies. To date, urban water networks, especially sewer systems, have been seen as potential sources for heat recovery [10,11], while cold recovery has been considered more recently [12,13]. In Paris, industrial applications of both heat and cold recovery have been built recently [14–16]. However, the field performance of actual recovery systems has only rarely been evaluated. Furthermore, cold recovery has never been considered as a means of responding to heat-waves.

In this paper, the Paris metropolitan area is used as a case study for evaluating different possibilities of using water as heating and cooling alternatives. For the heating supply from sewage water heat recovery, the field performance of an existing heat recovery system is assessed. Regarding cold supply from water mains during extreme heat, a potential assessment is conducted for three cold recovery configurations designed as an emergency response to heat-waves.

The rest of the paper is organized as follows: first, the global urban water cycle is described with special attention to temperature level and thermal energy recovery potentials in each step. Then, annual running data from a waste water heat recovery project in Paris is analyzed. Greenhouse Gas (GHG) emissions and primary energy savings are used as evaluation criteria. The third part gives innovative concept descriptions of three cooling productions from potable or non-potable water mains. These concepts are expected to provide real active solutions during heat-waves in high density urban centers.

## 2 Thermal energy recovery in water networks

The overall water cycle in an urban area begins at a river or underground water source and ends at the outlet of waste water treatment plants (WWTP). As shown in **Figure 2**, after being pumped from the source, treated water is transported to its end-users through urban water mains. After being used, sewage is carried to a WWTP via the sewer network. Certain cities, such as Paris, are equipped with secondary water networks dedicated to non-potable uses such as green space irrigation or street cleaning. This water may also come from a similar water source with less intensive treatment or may also be treated waste water, the source being the WWTP outlet. Regardless of the specifics, its cycle remains similar to that depicted in **Figure 2**.

Considering the whole urban water cycle, domestic hot water (DHW) preparation is by far the highest energy consumer, representing approximately 85 % of total energy needs [17]. The other two main energy uses are found at the supply and sewer disposal ends of the cycle. As a means of comparison, raising water temperature by 1°C is already equivalent to the energy needs of those two processes. Generally, DHW is heated to 60-65˚C to combat bacterial hazards, particularly *Legionella spp.* Given that the water inlet is between 10° and 15°C [18], the temperature must be raised by 45-55°C on average throughout the year, not accounting for seasonal variations.

Temperature levels in the whole water cycle range from 1°C to 65°C, as shown in **Figure 2**. In the cycle, two thermal energy recovery potentials can be possible: cold recovery in the water mains where temperatures are below 25°C, as well as heat recovery in the sewer systems where temperatures are between 13 and 35°C.

For heat recovery, the sewer water must remain above 13°C to meet the operational needs of WWTP processes. For cold recovery, water mains temperature must remain below 25°C to ensure that bacterial growth remains limited [19]. Therefore, in the case of closed loop systems (sewage or potable water), where water remains in the water network, a maximum temperature difference is permitted. However, in the case of an open loop system (introduced in section 4.3), higher temperature changes are allowed.

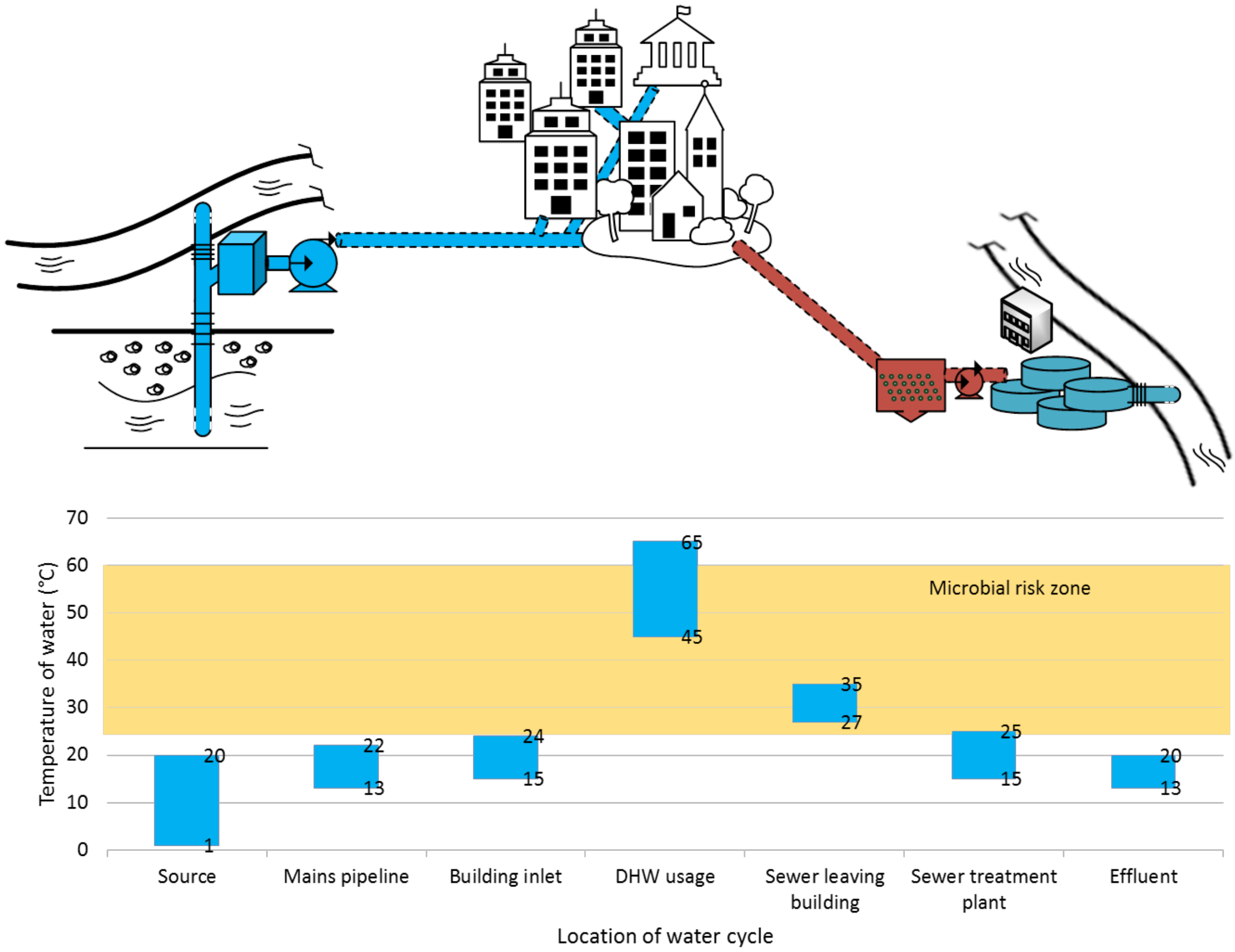

**Figure 2.** Temperature distribution in water cycle, from drinking water to sewage water

While the flow rate fluctuation feature of water networks could be a difficulty for recovery projects, its application at the district level is less intermittent. As long as the connecting population is sufficiently dense and diverse, a continuous flow rate is maintained almost all through the day. Particularly in the case of waste water heat recovery, sewer networks can temporally hold high effluent inlets since their volumes are in generally over dimensioned. Consequently, they can serve as buffers to stable waste water flowrate. In this paper, we focus our attention to the collective utilisation of water thermal resource, i.e., by supposing stable flow rates during heat recovery processes.

# 3 Heat recovery from sewage water system

## *3.1 Principle*

Waste water effluent has a temperature range of 35-27˚C at the outlet of buildings. In France, the temperature level decreases along sewerage channels until 13˚C before entering WWTPs. Lower temperatures should be avoided as most treatment processes require a warm environment for efficient nitrogen removal [20], even though recent studies have proposed low-temperature treatments [21,22]. From a heat recovery point of view, we consider a temperature level from 35˚C to 13˚C in this study.

Though temperature levels are higher closer to building sewer outlets, flow rates are lower and intermittent. Therefore, for district-scale systems heat recovery is only feasible in sewer collectors where the flow is continuous, despite lower temperature levels. Closer to the effluent source where temperatures are higher, individual solutions compatible with intermittent flows can be used for heat recovery for DHW systems for example [10,23].

At the district-scale, many centralized heat recovery projects exist and can be referred to. In such systems, a heat transfer fluid (HTF) is used to transport the thermal energy from the waste water to a heat pump (HP) water heater. One advantage of district-scale recovery is the possibility of providing DHW in addition to heating. By doing this, the payback period is shorter and the running hours of the installed equipment are greater. An example of a central heat recovery system is depicted in **Figure 3**. In such a system, two major components should be addressed: heat exchangers and HP.

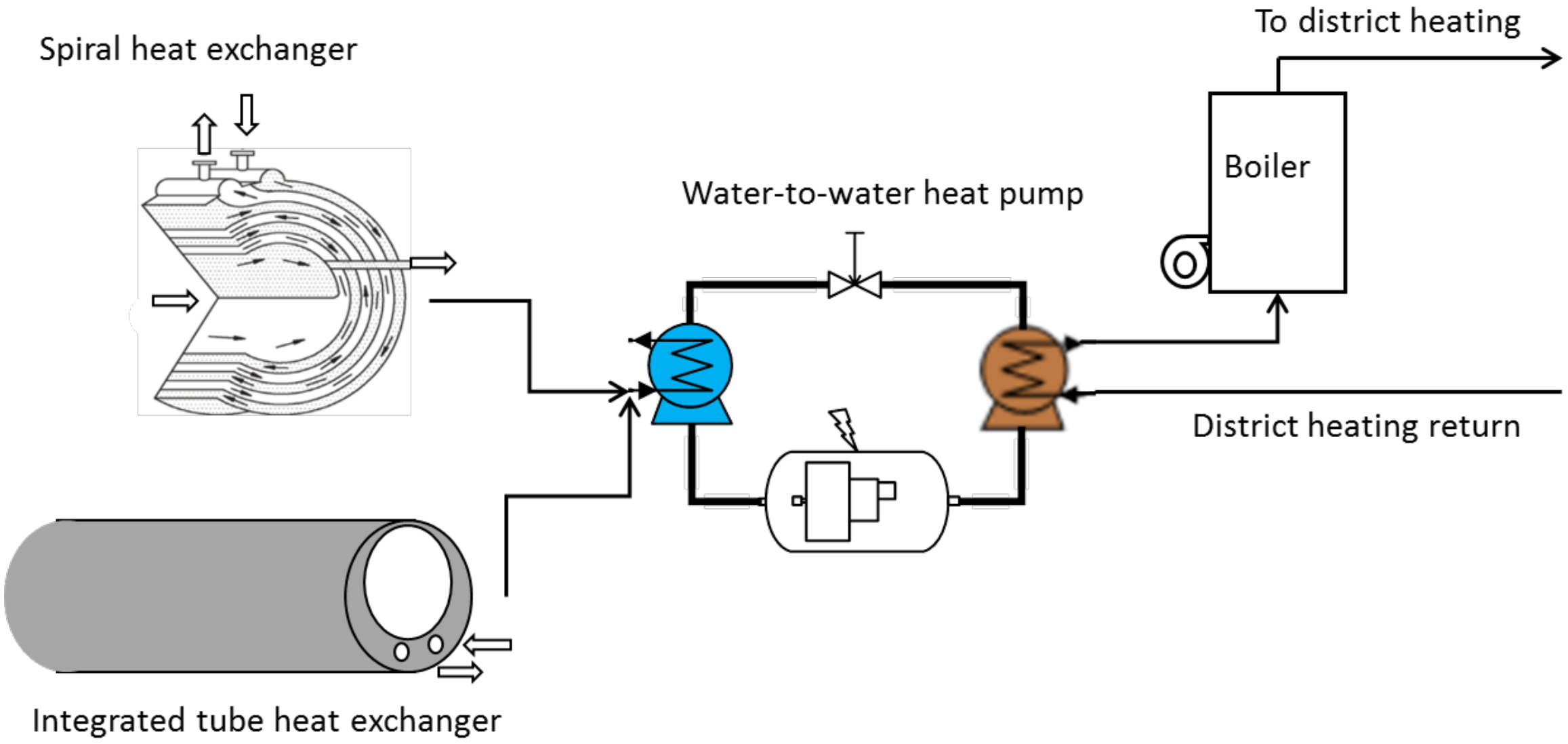

**Figure 3.** District heating with sewage water heat recovery integration

### *3.2 Heat exchanger technology*

At the district scale, two types of heat exchangers can be used: those incorporated into the sewage mains (integrated type) or spiral heat exchangers (external type). Both reduce the risk of fouling and frequent maintenance. Heat exchangers have three main characteristics: capacity, exchange surface and heat transfer coefficients. To minimize the temperature difference in waste water heat recovery, where a HTF is used, it is preferable to have a large exchange surface ($A$) and a high transfer coefficient ($U$).

Integrated exchangers can be added to the existing sewer system or incorporated into new sewer mains sections. The desired capacity is reached by assembling several modules or sections in series. Because of their comparatively low heat exchange coefficient, i.e. approx. 300 $W.m^{-2}.K^{-1}$ versus typical values of 2 000-5 000 $W.m^{-2}.K^{-1}$ [24], reaching the desired heat capacity often requires long segments. Spiral heat exchangers are much more compact thanks to their high exchange surface and high performance [25], but require a dedicated sewage network derivation. The choice between both technologies will tend to be dictated by the opportunity of adding a heat exchanger to an existing sewer system as opposed to creating a sewage derivation. Moreover, modern spiral heat exchangers hold self-cleaning anti-fouling options that reduce the frequency of maintenance. Integrated tube heat exchangers, however, should be regularly cleaned if the carried waste water is highly organic loaded.

Recent examples of both technologies can be found in France [14,26–29].

### 3.3 Heat pumps

Two important characteristics of HP should be considered: heating capacity and COP (Coefficient of Performance). Today, a large range of heating capacities are available, with nominal values from 2 kW to 20 000 kW [30]. Their production can be adapted to demand but only within certain limits: below a certain threshold, e.g. 10% of nominal capacity, the HP must be stopped and replaced by auxiliary heating devices such as a boiler. This vulnerability makes it important to consider the temporal fluctuations of both demand- and source-sides.

The COP of a HP is expressed as the ratio of its heating capacity and the compressor's electric consumption. In the literature, the COP for sewer heat recovery projects is reported to range from 1.8 to 10.6 [31].

The large performance range is partially due to the temperature differences between the hot and cold side, which is given by Carnot HP performance coefficient:

$$COP_{ideal} = \frac{T_h}{T_h - T_c} \quad (1)$$

In practice, due to exergy losses in a real cycle, we put about half of the Carnot COP value as that of real one, in addition to a part load factor considered:

$$COP_{real} = \frac{\dot{Q}_h}{W_{comp}} = 0.55 \times PLF \times \frac{T_h}{T_h - T_c} \quad (2)$$

where *PLF* stands for the part-load factor, and its value is lower than 1 when the HP runs at non-nominal conditions. $\dot{Q}_h$ and $W_{comp}$ represent respectively heating rate at the hot side and compressor electric power consumption.

For annual analyses, the Energy Factor (EF) is often used, which considers both compressor and auxiliary equipment consumptions as well as their seasonal variation. While temperature levels are relatively stable in our case (compared with air-source HPs with high seasonal variation), PLF can plays a major role in the final annual performance. The annual EF is given by:

$$EF_{annual} = \frac{\int \dot{Q}_h}{\int (W_{comp} + W_{aux})} \quad (3)$$

### 3.4 Case study

During the construction of a low carbon district in Nanterre (suburb of Paris), a 800 m long district heating network was built and is supplied by sewage heat recovery with an auxiliary gas boiler. The district is in a dense urban area and is transforming from a shut-down factory site to an eco-district. The transformation started in 2005 and the buildings are commissioned progressively between 2011 and 2017. In 2015, the mini-district heating network delivers heat and DHW to 650 high energy performance residential flats and a newly constructed school, totalling a heating surface of around 54 000 m$^2$.

The district heating system consists of two supplies providing respectively DHW and central heating and one common return. For the recovery side, an overall length of 200 m of heat exchanger is integrated into the sewer network, representing a total heat exchange surface of 112 m$^2$. The sewer network transports effluents at a flow rate of 115 m$^3$/h, rejected by around 15 000 equivalent habitants. The designed heat exchange capacity is 370 kW for a nominal temperature difference of 11 °C. The global heat transfer coefficient between HTF and waste water is 300 $W.m^{-2}.K^{-1}$. To raise the temperature level to the delivering value, i.e., 65 °C for DHW and 72 °C for district heating supply, two HPs are used. Each of them has a heating rate of 400 kWth. One of them is operated only for summer DHW supply and the other for heating and DHW in winter. The above parameters about this case study can be summarized in Table 1.

**Table 1.** Technical details of sewage heat recovery case study for the heating supply of an eco-district in Paris suburb

| **Thermal energy demands** | |
|---|---|
| Total heating ground ($m^2$) | 54 000 |
| Annual HDD in 2015 (site recorded, ˚C) | 1 670 |
| Annual heating demand (MWh) | 2 548 |
| DHW energy demand (MWh) | 1 207 |
| Heat delivery temperature (°C) | 72 |
| **Heat recovery by heat pump** | |
| Heat pump heating capacity (kWth) | 400 |
| Annual HP energy factor (-) | 2.7 |
| Electricity input to compression and auxiliary (kWM) | 1 417 |
| **Sewage water** | |
| HTF outlet temperature (˚C) | 10 |
| Heat exchanger capacity (kW) | 370 |
| Total heat exchanger surface ($m^2$) | 112 |

A yearly assessment has been conducted and monthly details are shown in **Figure 4** a/b/c. The total annual energy demand in 2015 (January to December) was 3 889 MWh with 2 548 MWh dedicated to space heating (October to May) and 1 207 MWh for DHW. 134 MWh were lost during the caloric transport. It is worth noting that the first semester of heating season in 2015 was particularly harsh, with monthly HDD of up to 349°C.day. The total annual HDD value of the year is 1 670°C.day. The main demand is central heating with an annual primary fraction of 66 %.

Total HP production supplies 84% of the annual energy demand, with the remaining 16% provided by gas boilers. The annual share of recovered energy (evaporator side of HP, shown in orange in **Figure 4**b) is 58 %, which makes the project eligible as a low carbon energy system for certain subsidies [32]. EF ranges from 2.6 to 3.0, except in July and August when the HP is only partially run due to low energy demand. A total of 1 165 MWh of electricity is consumed by the district heating through HP compressors, representing 26% of the total thermal energy demand. Another 252 MWh of electric energy is consumed by auxiliaries, i.e., pumps and flow distribution.

Operation maintenances are carried out during the first week of May, as well as all through the summer period between late June and the end of September. During this period, the heat exchanger is entirely cleaned to avoid fouling effect to the system performance.

**Figure 4**c shows monthly HP production as a share of total primary energy consumption. In France, electric energy must be multiplied by 2.58 for conversion to primary energy, while recovered heat is already considered to be primary energy. **Figure 4**c exposes the "dark side" of HP thermal energy recovery projects: while HP production satisfies at least 75% of the total demand during most months (shown as ○ in the figure), the recovered energy only represents 30-40% of monthly primary energy consumption (shown as green bars). The remainder, i.e. gas and electricity, represents 68% of primary energy consumption over the year. This issue is particularly embarrassing for HP recovery projects since building energy regulations generally consider primary energy consumption, not final energy. In consequence, compared with a gas boiler, a HP system whose EF is under 2.58 is not able to bring primary energy savings, even with waste heat recovery.

Finally, from a carbon emissions point of view, the waste heat recovery project is very promising. In France, GHG emission factors for electricity are estimated to be 0.055 t-eq $CO_2$/MWh and 0.206 t-eq $CO_2$/MWh for natural gas [33]. In addition, the gas boiler is considered to have an energy efficiency of 1.02 thanks to condensation generation. Results show that the case project has an annual carbon footprint of 204 t-eq $CO_2$ in 2015, including 127 t-eq $CO_2$ from the gas boiler, 64 t-eq $CO_2$ from the HP compressors and 14 t-eq $CO_2$ from auxiliaries. With a gas-only heating supply, the GHG emissions would have been 799 t-eq $CO_2$, i.e. almost 4 times more. This is particularly interesting in the light of the recent Climate Plan being developed by the French government which has recently set the goal of carbon neutrality by 2050 [34].

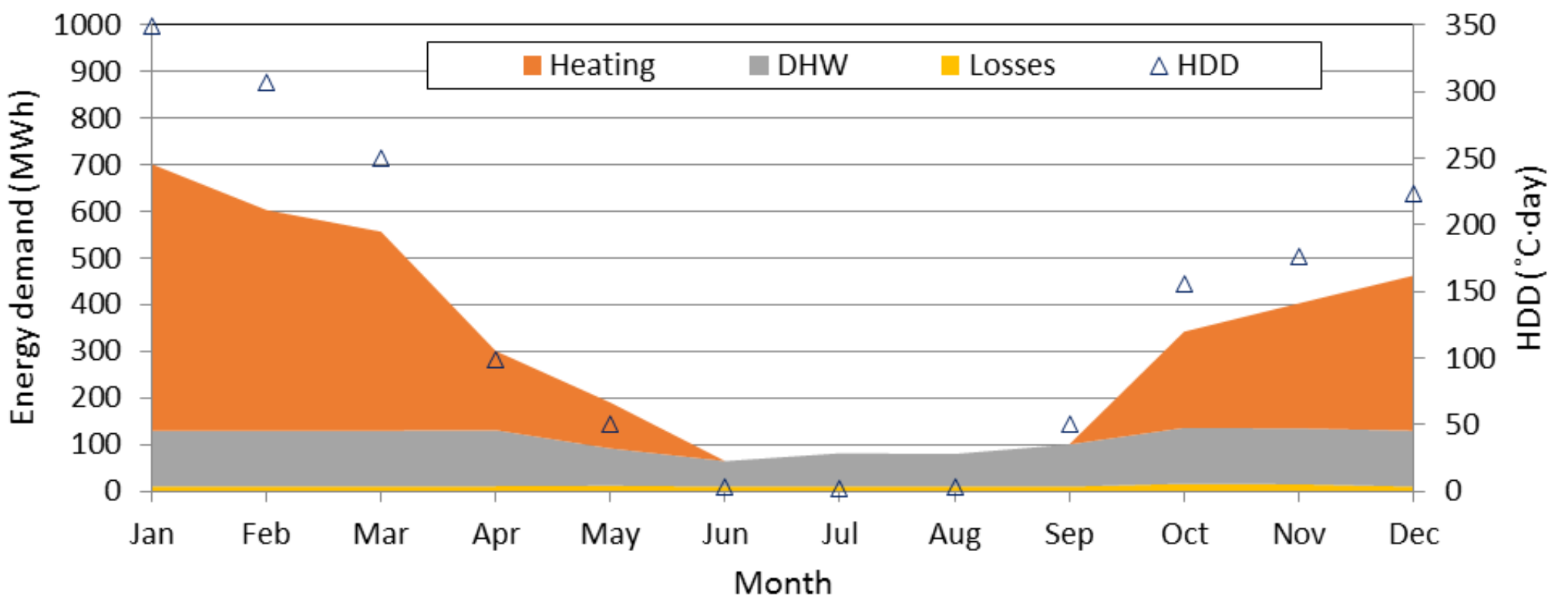

a) Energy demand from central heating, DHW and heat losses

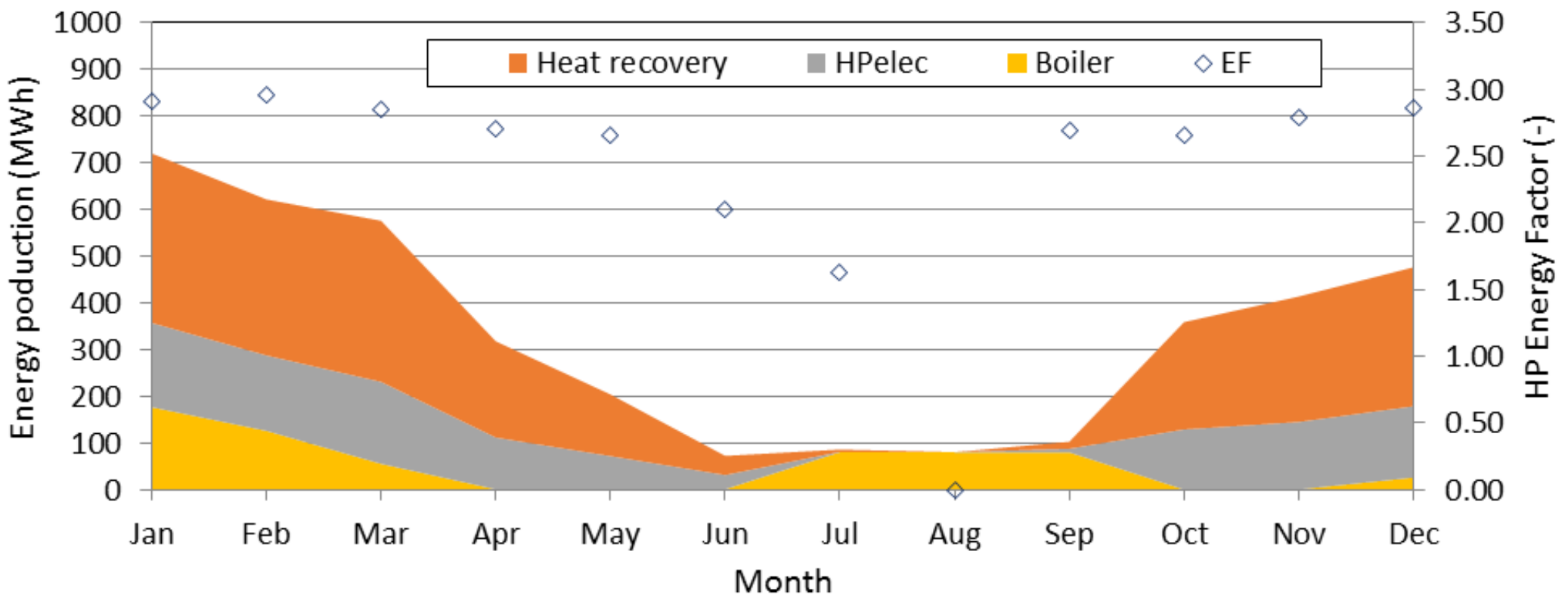

b) Production from waste water heat recovery, gas boiler and HP electricity

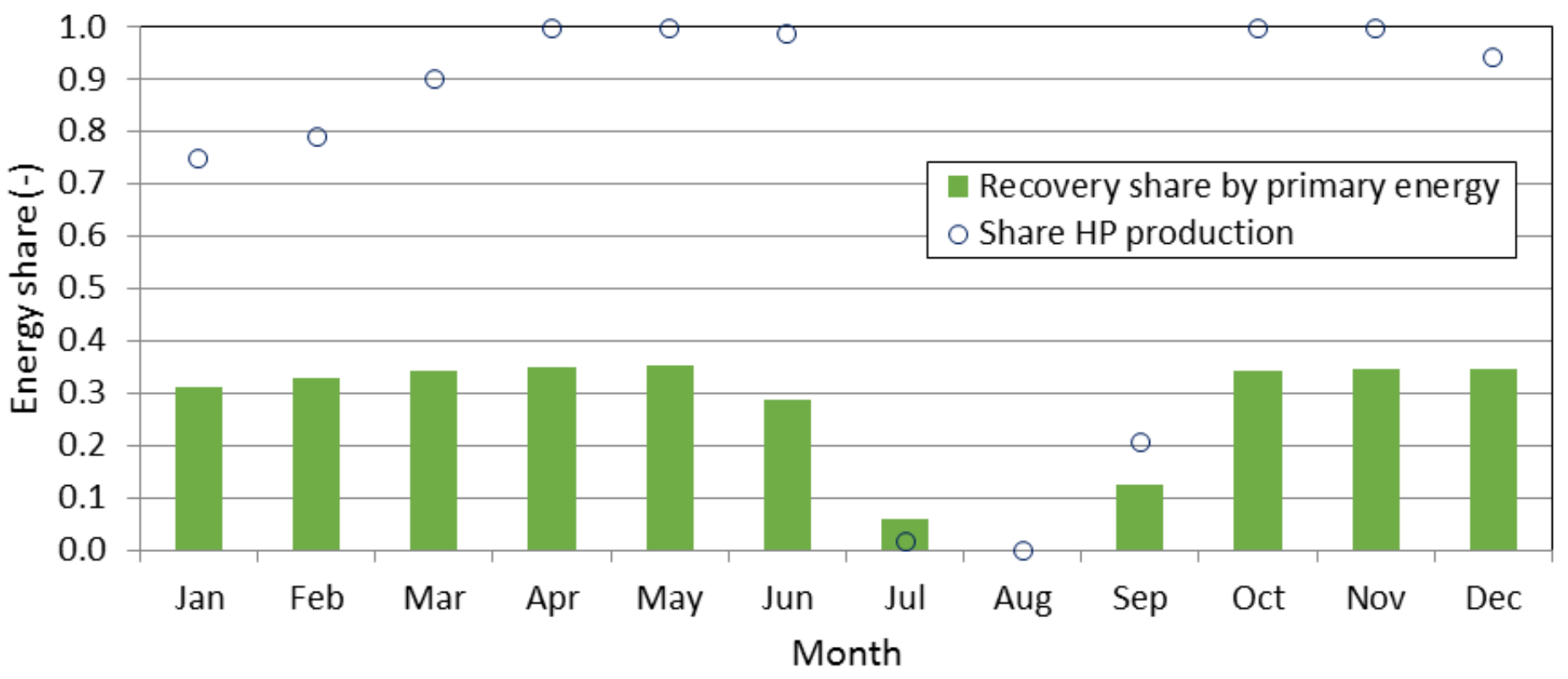

c) Total HP production and recovery share of heating demand, converted to primary energy

**Figure 4.** District heating with sewage water heat recovery performance data from on-site monitoring

# 4 Cold recovery from drinking or irrigation water networks

## *4.1 Underground railway station cooling*

In the Paris Metro system, stations are equipped with mechanical ventilation. The system is designed mainly against air pollution or for smoke extraction, but stations are not actively cooled. During heat-waves, high ambient temperatures and dense crowds can make these stations very uncomfortable and thermally stressful for passengers.

One way of containing temperature extremes is to use water mains cold recovery to provide cooling to underground railway stations. In **Figure 5**, we propose to transform existing air supply systems (2) into complete Air Handling Units (AHU) with an integrated heat exchanger. Thanks to a HTF, cooling produced by a refrigeration system (3) is delivered to the AHU heat exchangers. On the cold recovery side, another HTF provides condenser cooling and injects heat into the water mains through a heat exchanger (1).

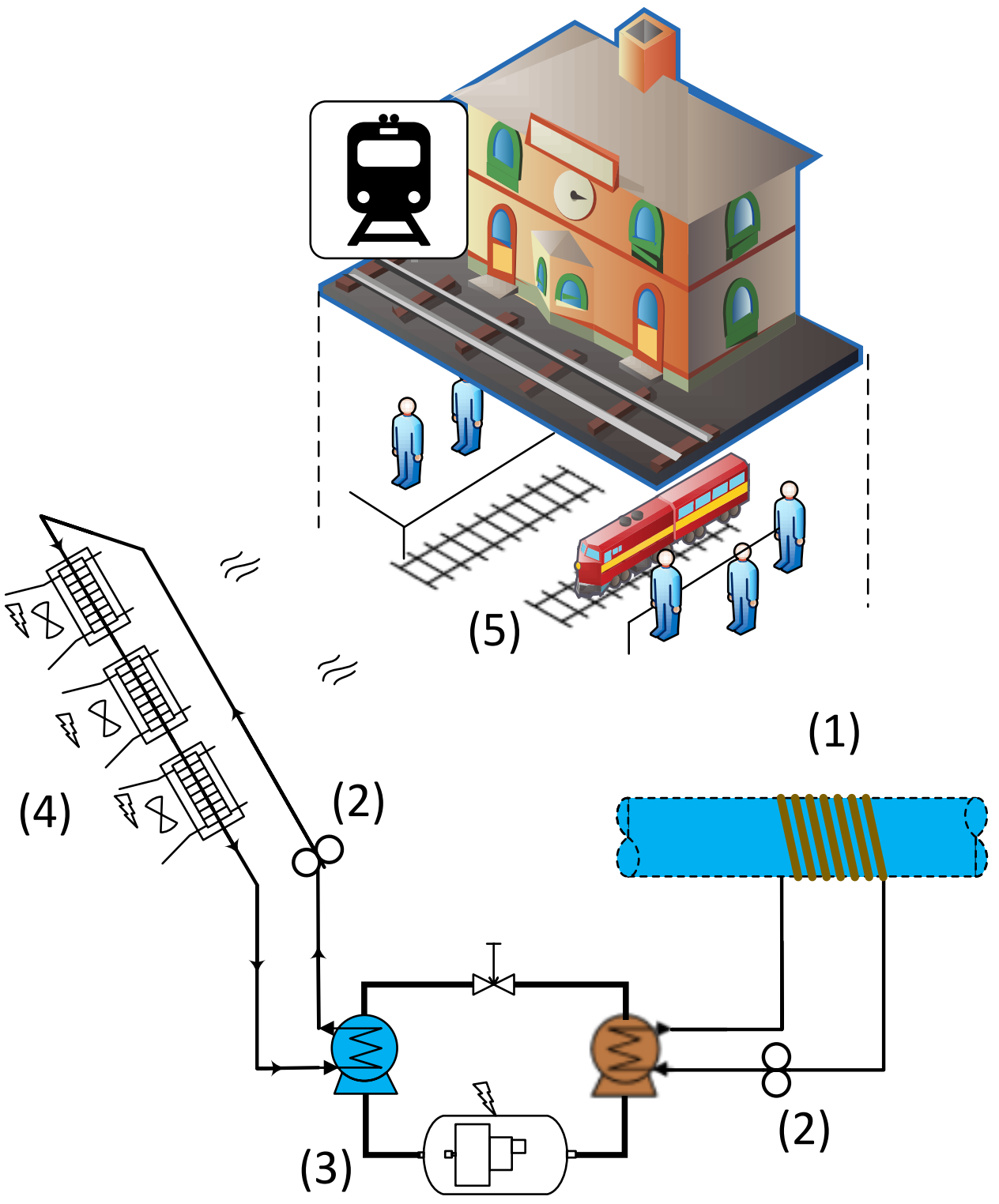

**Figure 5.** Metro station cooling by AHU supplied with cold recovery from water mains (1) heat exchanger, (2) circulator, (3) water-to-water refrigerator, (4) AHU, (5) underground Metro station

The following assessment is conducted for the Paris metro station Porte de Clignancourt, a typical small-sized metro station with only one passenger line (Line 4). While no active cooling is operated in the metro station, the configuration shown in **Figure 5** could help reduce air temperature by blowing cooled air (26°C) into the station instead of untreated outside air (35˚C).

Cooling the supply air in this manner would require 261 kW of refrigerating power. For a refrigeration unit with an energy efficiency ratio (EER) of 6.9 (e.g. Cooling unit Multistack [35], technical details in Table 2), 38 kW of electric power would be required. A total 299 kW of heat would then need to be dissipated into the water mains. For a 1.2˚C temperature increase, a mains flow rate of 216 $m^3$/h would be required. Supposing that the flow rate is continuous over 24h, this flow rate corresponds to the daily consumption of 25 920 people (~200 L per person per day). In Paris, the observed average water consumption per capita is approximately 230 L/day.

**Table 2.** Estimation details of water mains cold recovery to provide metro station cooling

| | |
|---|---|
| **Metro station load estimation** | |
| Air blowing flow rate ($m^3$/h) | 86 400 |
| Outside air temperature (˚C) | 35 |
| Blowing temperature (˚C) | 26 |
| Cooling load (kW) | 261 |
| **Cooling unit** | |
| Manufacturer/Model/Refrigerant | Multistack/MS050XN/R410a |
| Evaporator HTF outlet (˚C) | 16 |
| Condenser HTF inlet (˚C) | 24 |
| EER (-) | 6.9 |
| Electricity input (kW) | 38 |
| **Mains water** | |
| Cold recovery heat exchanger temperature difference (˚C) | 4 |
| Source-side heat exchanger capacity (kW) | 299 |
| Mains water inlet temperature (°C) | 20 |
| Potable water temperature increase (˚C) | 1.2 |

Approximatively 300 similar stations are distributed over Paris. Given the total flow rate of potable water, the described method could be deployed to 99 such stations located near potable water mains, provided that the initial water temperature permits it. This capacity could also be increased to 141 stations by also using the non-potable water mains, which supply an average 220 000 $m^3$/day.

It is worth mentioning that, due to fire safety regulations, most metro stations are equipped with air ventilation systems. Adding an air-water heat exchanger before existing ventilators can be done easily. This facilitates the implementation of metro station cooling.

### *4.2 Movable solution for night cooling during heat-waves*

At the individual scale, cooling is generally provided by an air-conditioning (AC) system. Few such units are installed in Paris, namely due to how short their use period would be, in addition to their aesthetic implications which may seem ill-suited to Paris' historic center. Moreover, the heat released by AC units outdoors intensifies the UHI effect and thus worsens the impact of heat-waves, particularly for pedestrians and persons not equipped with AC units. While there is significant improvement to be expected from behavioural adaptation to heat-waves [36], one alternative for active cooling, also based on closed loop mains water cold recovery, is to produce ice with a chiller unit and let people collect the ice to cool their bedrooms during particularly hot nights. Our following estimations are based on parameters of typical Paris buildings and a special attention is given to the ice quantity needed.

This mobile cooling solution is illustrated in **Figure 6**. An ice maker (3) is used to produce and store low temperature ice blocks (4) at -9 ˚C. People in the neighbourhood, in particular those identified beforehand as being vulnerable to heat-waves, e.g. those already signed up to the CHALEX list [37] used in Paris, collect or are delivered the necessary amount of ice in the evening. The ice is then allowed to melt (5) in their bedrooms during the night. A blower can be added to intensify the melting process by forced ventilation. The cold recovery by heat exchanger (1) is similar to that described in section 4.1.

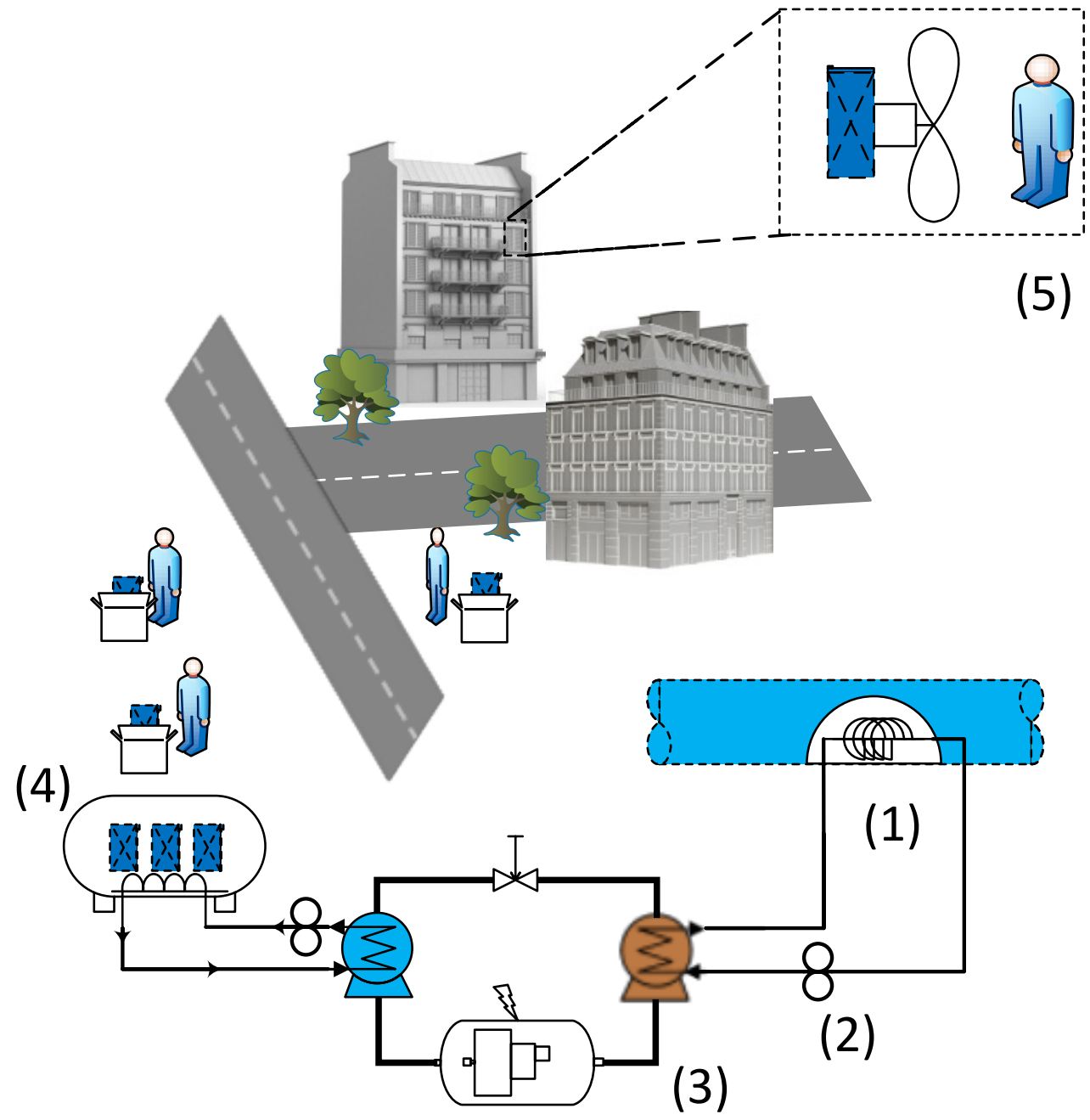

**Figure 6.** Night cooling by ice blocks produced by cold recovery from mains water (1) heat exchanger, (2) circulator, (3) water-to-water glace maker, (4) modular ice storage, (5) individual night cooling by movable ice

Load calculation and sizing are detailed in Table 3. On the load side, a standard 12 $m^2$ bedroom is occupied by two persons for 8h at night. The heat loss coefficient, i.e. the U-value of the 4 lateral walls, is 0.75 W/($m^2$·˚C). Cooling loads from floor and ceiling are considered negligible since they are next to adjacent bedrooms. The hourly air replacement rate is assumed to be 1 ACH (Air Change per Hour), with an outside temperature of 30˚C and a setpoint temperature of 26˚C in the bedroom. An internal load of 100 W is considered from the blower and other electronics, in addition to 120 W representing two sleeping occupants. In total, the cooling load is 394 W/bedroom, i.e. 3.15 kWh/bedroom/night.

Considering both sensible and latent energy, melting 1 kg of ice from -9˚C to 26˚C requires a total enthalpy of 0.128 kWh. Therefore, 24.7 kg of ice are required for one night, per bedroom. While not negligible, this mass is acceptable and could be delivered every day over several days under a heat-wave emergency situation.

Running under such conditions, a typical chiller runs at an EER of 1.4 [38]. This value is estimated by taking -13˚C as the evaporation temperature and 25˚C as the condensation one through equations (4)-(5). They provide respectively the ideal Carnot efficiency and practical value reduced by 4.

$$EER_{ideal} = \frac{T_h}{T_h - T_c} \quad (4)$$

$$EER_{real} = 0.25 \times \frac{T_c}{T_h - T_c} \quad (5)$$

If we consider a total of 1 200 bedrooms, and the total ice demand is 29 597 kg, equivalent to 3 786 kWh of cooling. If the chiller is operated continuously during 24h, the electricity consumption and condenser heat gains are 115 kWh and 273 kWh, respectively. Assuming the same flow rate as in section 4.1, the water temperature would increase by 1.1°C.

Scaled to the whole city's potable water network, nearly 120 000 bedrooms could be cooled this way, or up to 170 000 if the non-potable water network was also included. Given that there currently are 8 400 people on the CHALEX list in Paris [37], the potential for cooling their bedrooms greatly exceeds demand.

**Table 3.** Estimation details of water mains cold recovery for ice production

| | |
|---|---|
| **Bedroom cooling load estimation** | |
| Room dimensions (m) | w*l*h: 3*4*3 |
| Heat loss coefficient U, (W/(m2·˚C)) | 0.75 |
| Air renewal rate (cycle/h) | 1 |
| Internal load, 2 persons sleeping, (W) | 2x60 |
| Internal load, blower and other electronics, (W) | 100 |
| Outside air temperature (˚C) | 30 |
| Room temperature (˚C) | 26 |
| Cooling load (W) | 394 |
| Night duration (h) | 8 |
| Number of bedrooms (-) | 1 200 |
| Total chilling demand (kWh) | 3 786 |
| **Ice quantity** | |
| Initial temperature (˚C) | -9 |
| Final temperature after melting (˚C) | 26 |
| Total enthalpy initial-final (kWh/kg) | 0.128 |
| Quantity of ice for a bedroom during 8h (kg) | 24.7 |
| Quantity of ice for all bedrooms (kg) | 29 597 |
| **Chiller – ice maker** | |
| Manufacturer/Model/Refrigerant | Teknotherm/F-45/R404a |
| Evaporator HTF outlet (˚C) | -13 |
| Condenser HTF inlet (˚C) | 25 |
| EER Carnot (-) | 6.8 |
| EER (estimated, -) | 1.4 |
| Daily electricity consumption (kWh) | 2 767 |
| **Mains water** | |
| Cold recovery heat exchanger temperature difference (˚C) | 4 |
| Source-side heat exchanger capacity averaged by 24h (kW) | 273 |
| Potable water flow rate ($m^3$/h) | 216 |
| Equivalent number of inhabitants (-) | 25 920 |
| Mains water inlet temperature (°C) | 20 |
| Potable water temperature increase (˚C) | 1.1 |

### *4.3 Emergency urban cooling during heat-waves*

One response to heat-waves currently under consideration for cities such as Paris is pavement-watering [39,40]. By depositing a water film on street surfaces, the method provides evaporative cooling able to positively affect air temperature and pedestrian thermal stress [41]. Field measurements conducted in Paris have reported cooling of up to 0.8°C and 3.7°C in air and mean radiant temperatures with a maximum increase of relative humidity of 4%. These combined microclimatic impacts resulted in a positive impact on pedestrian thermal stress with a maximum reduction of the Universal Thermal Climate Index (UTCI) equivalent temperature of 1.5°C [42].

In Japan, pavement-watering systems are already installed in the streets of Nagaoka City for melting snowfall and have been used to study pavement-watering for summertime cooling [39]. In Paris, a similar pavement-watering infrastructure could be based on the city's non-potable water network, already present in most of its streets. **Figure 7** illustrates how such a system could be installed and connected to the existing water network. A heat exchanger for cold recovery could be placed between the sewer and pavement sprinkler. Since pavement-watering would only be activated during a heat-wave, the associated water use would coincide with the proposed emergency cooling systems discussed previously.

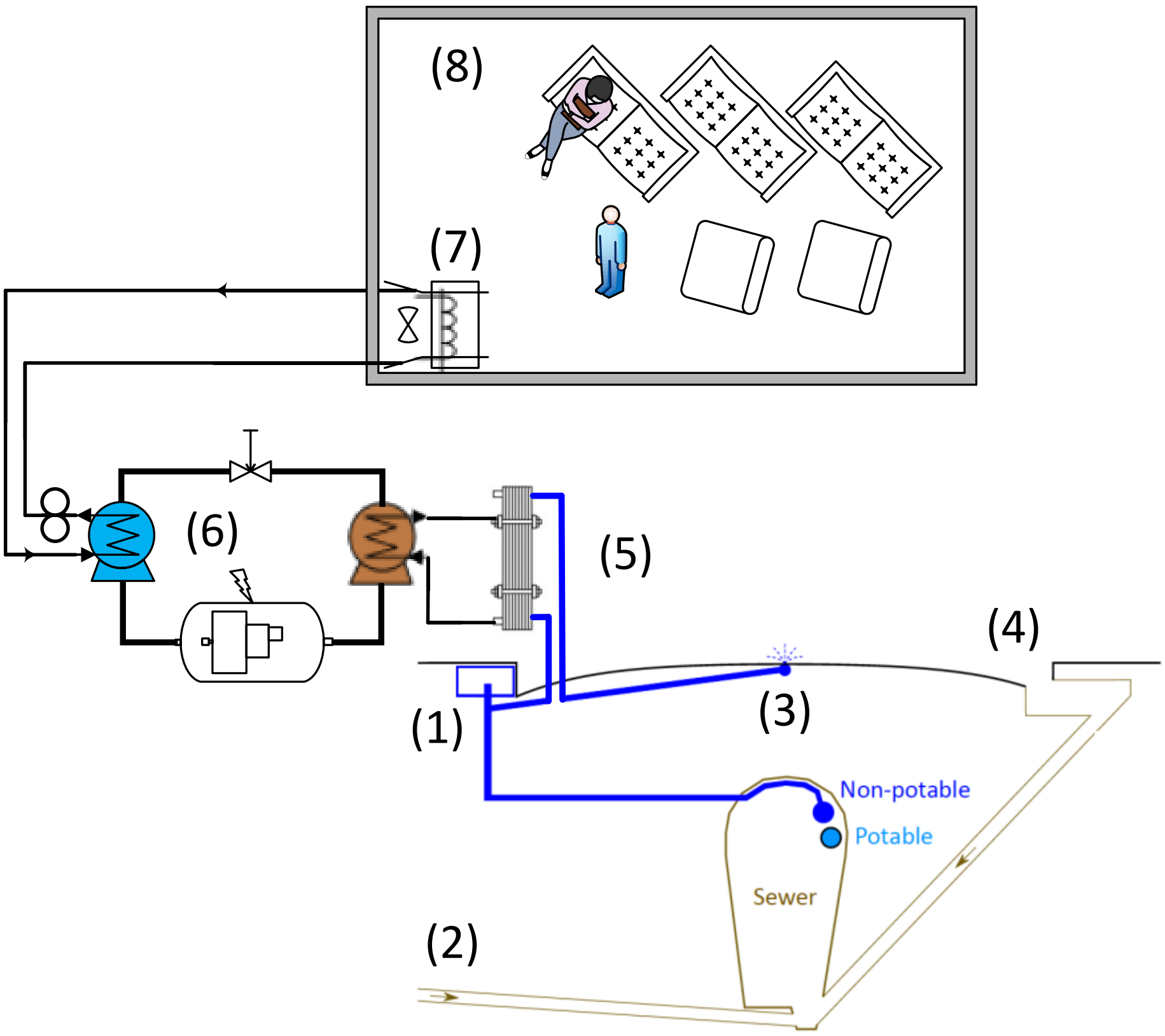

**Figure 7.** Cross section of street structure with sewer and water mains. The pavement-watering system could be connected to the non-potable water network.
(1) non-potable water supply, (2) waste water, (3) pavement sprinkler, (4) storm drain, (5) heat exchanger, (6) cooling unit, (7) AHU, (8) emergency heat-wave shelter

In their study, Hendel et al. analysed the thermal effects of pavement-watering and proposed an optimised watering strategy for their site [43]. Under direct insolation and with a watering rate of 0.41 mm/h, surface cooling of up to 265 W/m² was found, i.e. approximately 7 W/m² of sensible cooling and 257 W/m² of evaporative cooling. Sensible cooling therefore only accounts for 3% of total pavement-watering cooling.

Given its limited contribution, the sensible component could be used in a more concentrated form, e.g. for space cooling of nearby cafés, restaurants or public libraries, with only a minor impact on total pavement-watering cooling. To achieve this, a heat exchanger could be placed between the water main and the sprinkling nozzle in the center of the street in Figure **7**. The water being used for pavement-watering would simply be preheated from its initial temperature (20-25°C) to the water film temperature (35-40°C depending on the street's insolation conditions).

As opposed to the previous closed-loop approaches, where water remains or is reinjected into the network, this method is open-loop. While a closed loop requires low water temperature changes for there to be no impact on its quality and usability, an open loop allows a much higher temperature change to be used. Therefore, even small flow rates can provide large amounts of power.

As a case study, let us consider a 200 m long portion of street 20 m wide, i.e. an area of 4 000 m², watered at the rate of 0.41 mm/h reported in Hendel et al., i.e. approximately 20 $m^3$/day [43]. During daytime watering, the average delivered flow is 0.46 L/s. With a temperature gradient of 15°C, this flow can absorb 28.6 kW. Considering a heat pump COP of 3, approximately 21.4 kW of cooling power can be continuously produced in association with pavement-watering.

At the urban scale, pavement-watering would be conducted during a heat-wave in only the most intensively walked streets during direct insolation, located outside of the cooling reach of parks or rivers and where no shading is available to pedestrians. The cold recovery proposed here would therefore coincide spatially and temporally with the areas and periods identified as having the highest solar gain. However, while they are well suited for reaching the full potential of pavement-watering, these periods may not occur when temperatures or pedestrian activity are highest. *A priori*, cold recovery would target peak indoor temperatures which are reached around 8 pm or 9 pm (UTC+2) for Paris [36], i.e. after scheduled watering. To make up for this potential shortfall, thermal storage could be used to offset the recovered cooling energy to periods with higher temperatures and pedestrian traffic. A water tank prefilled the day previous to pavement-watering could be used for this purpose. This reservoir would allow for cold recovery independently of scheduled pavement-watering proportionally to its volume.

Furthermore, the proposed cooling could be used to create "heat-wave shelters" for pedestrians or residents living in particularly warm housing. These spaces could simply be part of an existing café, restaurant, public library or school or they could even be part of the street furniture, for example a cooled bus stop, telephone booth or something else designed for this purpose alone.

## 5 Conclusions and perspectives

The main contribution of this study is to demonstrate the effectiveness of using water as heating and cooling sources at the urban scale. To the authors' knowledge, very few studies in the literature assessed this possibility in a global point of view and in particular the cooling recovery potential. A heat recovery case study was conducted and three cold recovery solutions that might be used as part of an emergency heat-wave response strategy for Paris were presented and discussed.

Heat recovery from waste water is considered as a green alternative for heating and DHW production. The analyses conducted on our case study indicate GHG emission reductions of approximately 75%, from 799 t-eq $CO_2$ to 204 t-eq $CO_2$ with a HP system covering 84% of yearly demand. An annual primary energy saving of 32% is reached, limited by the HP system's EF which ranges from 2.6 to 3.0 over the year.

Three emergency heat-wave cooling strategies are explored, based on closed or open loop systems for potable or non-potable water cold recovery. Of the 300 metro stations present in Paris, 99 to 141 of them could be cooled by mains water cold recovery. Alternatively, up to 170 000 bedrooms could be cooled with ice blocks produced from water mains cold recovery. Both solutions result in a water temperature increase of around 1 ˚C, which is acceptable from a water quality standpoint. The last solution (open-loop), coupling higher temperature lifting (up to 20 ˚C) cold recovery from non-potable water used for pavement-watering, provides a spatially-distributed emergency cooling solution. For a 200 m long portion of road, the potential cooling production can reach 21.4 kW, able to cool a local heat-wave shelter.

Other sources of water are also available for heat or cold recovery in Paris or other cities. These include existing uses of potable or non-potable water, e.g. for street cleaning or green space irrigation which may be used similarly to the open loop system. Furthermore, ground water which seeps into underground structures such as metro stations or parking lots. Currently this water is mostly pumped directly into the sewer network without serving any thermal energy supply, despite stable temperatures throughout the year.

Heat pumps and heat exchangers are key elements for both heat and cold recovery. The performance of heat pump or refrigeration units depends highly on the temperature difference, and is thus closely related to heat exchange performance. The heat recovery case study is based on a temperature difference of 11 ˚C between waste water and the HTF. Efficient heat exchangers with high heat exchange coefficients can allow better heat pump performance and thus better primary energy savings. The spiral heat exchanger solution appears to be promising given its compact character. Otherwise, the integrated heat exchanger requires minimal maintenance but a large heat transfer surface (112 $m^2$ in the case study).

For cold recovery from drinking water, an adapted heat exchanger has yet to be developed. The main challenges they face include sanitary compatibility, compactness and low pressure drop. The possibility of pipeline integrated heat exchanger without direct potable water passage to heat exchanger is promising, but the heat exchanger coefficient should be higher than 300 $W \cdot °C^{-1} \cdot m^{-2}$ in order for the occupied length to be acceptable.

In the current estimation, the heat pump and chillers are simplified with manufacturer performances, and the heat exchanger configuration is not explored. Our future works will be concentrated on compact heat exchanger development as well as rigorous dynamic simulations considering demand- and source-side fluctuations.

**Acknowledgement**

The authors would like to acknowledge the contribution of Laurent Royon for his fruitful suggestions and discussions. This work is partially supported by EFFICACITY, a joint research institute for urban energy transition.